# EFFECT OF CAMERA'S FOCAL PLANE ARRAY FILL FACTOR ON DIGITAL IMAGE CORRELATION MEASUREMENT ACCURACY

Ala Hijazi[1,*], Nathir Rawashdeh[2], Christian Kähler[3]

**ABSTRACT**

The digital image correlation (DIC) method is one of the most widely used non-invasive full-field methods for deformation and strain measurements. It is currently being used in a very wide variety of applications including mechanical engineering, aerospace engineering, structural engineering, manufacturing engineering, material science, non-destructive testing, biomedical and life sciences. There are many factors that affect the DIC measurement accuracy where that includes; the selection of the correlation algorithm and parameters, the camera, the lens, the type and quality of the speckle pattern, the lightening conditions and surrounding environment. Several studies have addressed the different factors influencing the accuracy of DIC measurements and the sources of error. The camera's focal plane array (FPA) fill factor is one of the parameters for digital cameras, though it is not widely known and usually not reported in specs sheets. The fill factor of an imaging sensor is defined as the ratio of a pixel's light sensitive area to its total theoretical area. For some types of imaging sensors, the fill factor can theoretically reach 100%. However, for the types of imaging sensors typically used in most digital cameras used in DIC measurements, such as the "interline" charge coupled device CCD and the complementary metal oxide semiconductor (CMOS) imaging sensors, the fill factor is much less than 100%. It is generally believed that the lower fill factor may reduce the accuracy of photogrammetric measurements. But nevertheless, there are no studies addressing the effect of the imaging sensor's fill factor on DIC measurement accuracy. We report on research aiming to quantify the effect of fill factor on DIC measurements accuracy in terms of displacement error and strain error. We use rigid-body-translation experiments then numerically modify the recorded images to synthesize three different types of images with 1/4 of the original resolution. Each type of the synthesized images has different value of the fill factor; namely 100%, 50% and 25%. By performing DIC analysis with the same parameters on the three different types of synthesized images, the effect of fill factor on measurement accuracy may be realized. Our results show that the FPA's fill factor can have a significant effect on the accuracy of DIC measurements. This effect is clearly dependent on the type and characteristics of the speckle pattern. The fill factor has a clear effect on measurement error for low contrast speckle patterns and for high contrast speckle patterns (black dots on white background) with small dot size (3 pixels dot diameter). However, when the dot size is large enough (about 7 pixels dot diameter), the fill factor has very minor effect on measurement error. In addition, the results also show that the effect of the fill factor is also dependent on the magnitude of translation between images. For instance, the increase in measurement error resulting from low fill factor can be more significant for subpixel translations than large translations of several pixels.

**INTRODUCTION**

Digital image correlation (DIC) is one of the most widely used photogrammetric measurements methods. The DIC method has been successfully utilized in a very wide variety of applications including mechanical, aerospace, structural, manufacturing, material science, non-destructive testing, biomedical and life sciences, and much more [1-9]. The DIC method is capable of obtaining full-field measurements of the displacements and strains for the surface being observed. The idea of the DIC method is quite simple where the surface of interest is covered with a random speckle pattern and a camera records images of the surface while it undergoes deformation. The reference image (the first image in the sequence) is then divided into square groups of pixels (called subsets) and these subsets are digitally correlated with the corresponding subsets in the images acquired at different stages during deformation. As such, the displacement/deformation maps are obtained, and then the strain maps can also be obtained. There is a variety of factors that affect the accuracy of DIC measurements. The main factors influencing the DIC measurement accuracy include; the selection of the correlation algorithm and parameters (subset and step sizes, correlation and shape functions, sub-pixel algorithm, etc.), the camera (resolution, dynamic range, noise, etc.), the lens (optical aberrations and image quality), the type and quality of the speckle pattern, the lighting conditions and surrounding environment. There are numerus studies addressing the different factors affecting the accuracy of DIC measurements and the sources of error [10-16]. In a previous study, Hijazi *et al.* [12] reported that the DIC strain error is dependent on the direction of translation (i.e., whether it is in the horizontal or vertical direction of the image plane). They indicated that this directionality is likely to be attributed to the difference in the fill factor in the horizontal and vertical directions of the image sensor. Also, it is generally believed that the use of cameras with low fill factors can lead to lower accuracy in photogrammetric measurements [17, 18]. There are some studies addressing the effect of fill factor on resolution for infrared (IR) imaging sensors; however, there are no studies addressing the effect of the imaging sensor fill factor on photogrammetric measurements in general and DIC measurement accuracy in particular. The lack of such studies might be attributed to the fact that there is no direct and straightforward method for studying the effect of fill factor. In order to evaluate the effect of fill factor on DIC measurement accuracy experimentally, it is generally required to test different cameras having identical specs and performance parameters except for the fill factor. Finding such cameras is very difficult, if not impossible, and this makes such approach unfeasible.

[1] Department of Mechanical Engineering, German Jordanian University, Amman, Jordan
[2] College of Computing, Michigan Technological University, Houghton, Michigan, USA
[3] Institute of Fluid Mechanics and Aerodynamics, Munich Bundeswehr University, Munich, Germany
*E-mail address: ala.hijazi@gju.edu.jo

We report on research aiming to quantitatively study the effect of fill factor on DIC measurements accuracy. In order to do so, we use rigid-body-translation experiments then numerically modify the recorded images to synthesize reduced resolution images with different fill factors. Our approach uses row and column averaging for groups of adjacent pixels (i.e., binning) along with selective row/column deletion (i.e., subsampling) in order to generate images with different simulated fill factors. Three types of synthesized images are produced out of each of the original images: i) 100% fill factor, ii) 50% fill factor (50% in horizontal direction and 100% in vertical direction), and iii) 25% fill factor (50% in both the horizontal and vertical directions).

**BACKGROUND**

A digital imaging device (or camera) consists of an imaging optic (a lens) which collects the light emanating from a target, and projects a real image of that target on an electronic image sensor. The electronic image sensor, which is usually referred to as the focal plane array (FPA), consists of a matrix of capacitor-like storage elements, known as pixels, formed on an oxide covered silicon substrate. The silicon has photoelectric property where it converts the incident light to electrical charge. As an optical image is projected on the FPA, the photons reaching each pixel generate an electrical charge having a magnitude that is proportional to the local intensity of light falling on that pixel. After exposing the FPA to light for some period of time, known as the exposure time, a pattern of charges is collected in the pixels (meaning that an image has been captured in the FPA). This pattern of charges is then transferred out from the FPA and read-out to a data storage device (after being digitized), thus freeing the FPA to capture another image. The two main types of electronic imaging sensors are the charge-coupled device (CCD) and the complementary metal oxide semiconductor (CMOS). Although both the CCD and CMOS sensors were invented around the same time, the CCD technology developed much faster and became more dominant (due to the more complicated design of CMOS sensors). However, nowadays, the use of CMOS sensors has become more widespread and they are replacing CCD sensors in many imaging applications. The main difference between CCD and CMOS sensors is in the way an image (i.e., the pattern of charges collected in the pixels) is transferred out of the pixels after it has been captured. For the interested reader, a detailed review of the different types of electronic imaging sensors can be found in references 17 and 21.

The fill factor is one of the important parameters for the FPA of a digital camera, though it is not widely known and usually not reported in specification sheets. The fill factor of an imaging sensor is defined as the ratio of the FPA's light sensitive area to its total theoretical sensory area. The fill factor is usually given as a percentage and it can theoretically reach 100% for some types of imaging sensors. The higher the fill factor, the more sensitive the imaging sensor is to light, increasing its quantum efficiency. The three most commonly known types of CCD sensors are the "full frame" CCD, the "frame transfer" CCD, and the "interline" CCD [17, 21]. Figure 1 shows the layout of the full frame CCD and the interline CCD. The full frame CCD, figure 1 (a), has a theoretical fill factor value of 100% since the entire area of the FPA is light sensitive. However, with this type of imaging sensors, an external shutter must be used in order to shield the FPA from light while the image captured in the FPA is being readout sequentially, which takes a relatively long time. The interline CCD, figure 1 (b), has light sensitive image sections and storage sections, which are shielded from light. There, the image and storage sections are interlaced as seen in the figure, either in the form of interlaced columns or interlaced rows. In interline CCDs, each pixel has its corresponding storage section where the charge in each pixel is quickly transferred to the storage section after the end of the exposure time. This configuration has the advantage of eliminating the need for an external shutter; however, this comes at the expense of reducing the fill factor. The same is true for CMOS imaging sensors where the fill factor will be even lower than that of interline CCDs since each pixel will have charge to voltage conversion, amplification, and digitization circuits onboard. The fill factor for some types of high speed imaging sensors can reach as low as 13% [17]. For the types of imaging sensors typically used in most digital cameras used in DIC measurements, such as the interline CCD and the CMOS imaging sensors, the fill factor is roughly in the range of 40% to 70%. In order to overcome the issue of low fill factor and achieve high quantum efficiency, many of the modern high-end imaging sensors (typically used in scientific imaging applications) utilize on-chip micro-lens arrays [21]. The use of micro-lenses for each of the individual pixels of the FPA will significantly increase the fill factor where it can practically approach 100% and thus the issues associated with low fill factor are eliminated. It should be noted here that the fill factor is calculated as an area ratio; however, a linear fill factor can also be defined for each of the horizontal and vertical directions of the FPA. If we refer to the interline CCD shown in figure 1 (b) for instance, it can be seen that the linear fill factor in the vertical direction is basically 100% while the linear fill factor in the horizontal direction is about 60%, which is also equal to the overall fill factor of the imaging sensor. The same is also true for many types of CMOS chips where the linear fill factors in the horizontal and vertical directions have different values.

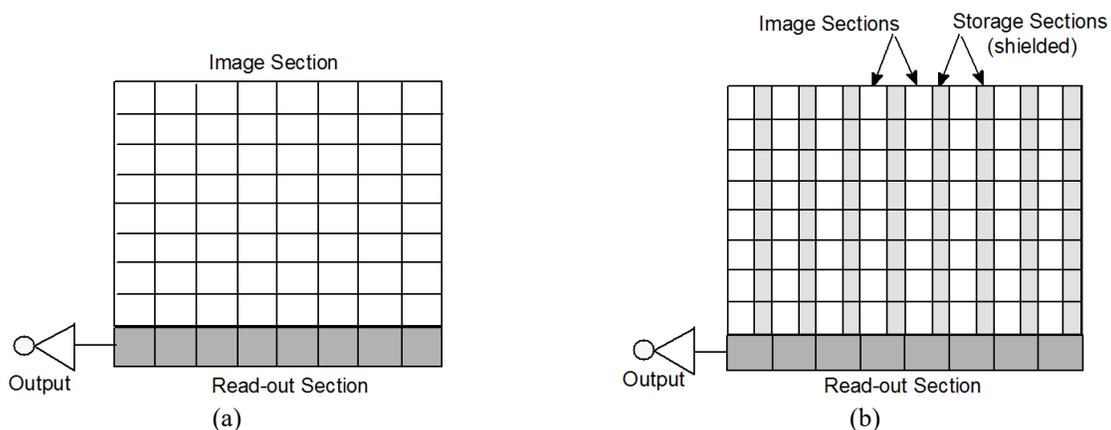

**Figure 1.** Different types of CCD imaging sensors; a) full frame CCD, b) interline CCD.

## APPROACH
*Experimental Setup*

The imaging system used for performing the experiments consists of a 5.5 megapixel monochrome scientific imaging camera (PCO Edge 5.5) fitted with a premium quality 50mm focal length lens (ZEISS Milvus 2/50M). The PCO Edge 5.5 camera has an 18.8 mm size Scientific CMOS (sCMOS) chip with 2560×2160 pixels resolution and 16-bit dynamic range. The sCMOS chips are fitted with micro-lens arrays, where that increase the fill factor and thus quantum efficiency (QE) to over 60%. The camera is fixed on an adjustable multi-axis camera-mount (to allow adjustment of camera orientation) which is in tern mounted on an optical rail such that the camera can be brought to the desired working distance. The target plate is mounted on a multi-axis high-precision translating/rotating stage. An overall view of the experimental setup is shown in figure 2. Before starting the experiments, the camera is brought in contact with the target surface such that the camera orientation is adjusted to ensure that it is perfectly perpendicular to the target surface. Then, the camera is retracted back to the appropriate working distance such that the field-of-view observed by the camera is 100 mm wide (that makes the scale factor to be about 26 pixels/mm). Printed random speckle patterns are used in the experiments where they are affixed on the surface of the target plate using two-sided tape. Three different random speckle patterns are used in the experiments and they are shown in figure 3. Speckles (a) and (b) are high contrast speckle patterns (black dots on white background) but with different dot sizes, where (a) has 0.5 mm dot diameter and average dot spacing of 0.75 mm; while (b) has 0.24 mm dot diameter and 0.36 mm average spacing. Speckle pattern (c) has relatively lower contrast compared to (a) and (b). It was produced by taking a defocused image of speckle (a) and printing it out. Using each of the three speckle patterns, two images are recorded by the camera; a reference image and another image after translating the target 5 mm along the *x*-axis (i.e., in the horizontal direction).

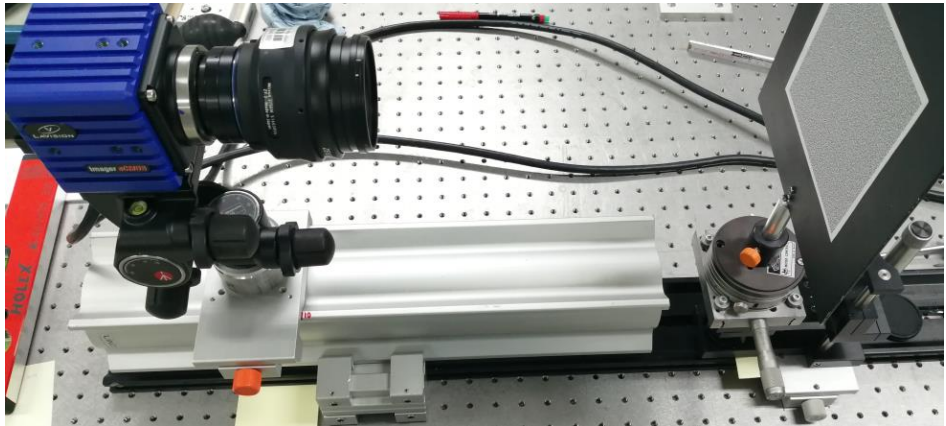

**Figure 2.** Setup used for rigid-body-translation experiments.

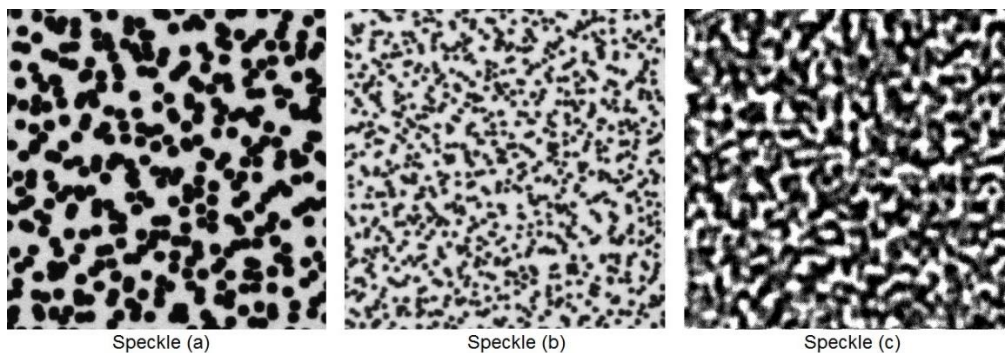

**Figure 3.** Close-up view of the different speckle patterns used in the experiments.

*Fill Factor Modification*

After obtaining the images recorded by the camera, these images where numerically manipulated to produce different sets of images representing different fill factors. It should be noted here that the images recorded by the camera are considered to have a 100% fill factor especially since the camera's sCMOS chip is fitted with a micro-lens array. A specially developed MATLAB code is used to generate three reduced resolution images (1/4 of the original resolution, i.e., 1280×1080) from each of the original images. The three generated images represent different values of the fill factor, namely; 100%, 50%, and 25%. The different schemes used for producing each of the three different types of images are illustrated graphically in figure 4. The images repressing 100% fill factor are produce by performing 2×2 binning (i.e., averaging 2×2 pixels cells) as illustrated in figure 4 (a). The images repressing 50% fill factor are produce by performing binning for the rows and subsampling for the columns (i.e., averaging even and odd rows, and deleting even columns) as illustrated in figure 4 (b). It should be noted that the 50% fill factor images has a linear fill factor of 50% in the horizontal direction while the linear fill factor is 100% in the vertical direction. Finally, the images representing 25% fill factor are produce by performing subsampling for both rows and columns (i.e., deleting even rows and columns) as illustrated in figure 4 (c). It also should be noted that for the 25% fill factor images, the linear fill factor is 50% in both the horizontal and vertical directions.

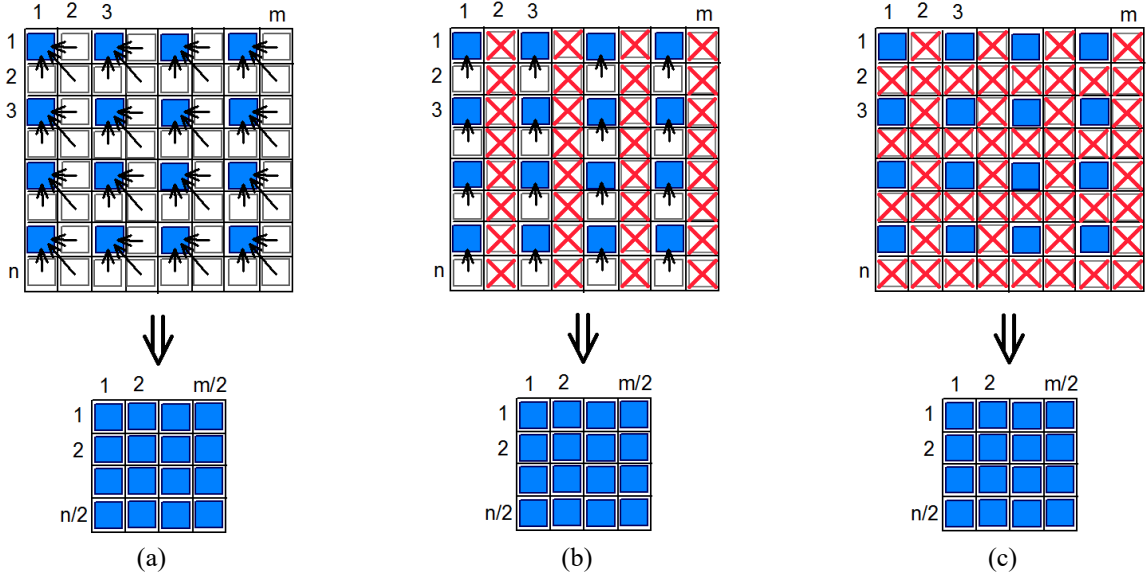

**Figure 4.** Image manipulation to produce reduced resolution images with different fill factors; a) 100% fill factor, b) 50% fill factor (50% horizontal × 100% vertical), and c) 25% fill factor (50% horizontal × 50% vertical).

*DIC Analysis*

For each of the three different speckle patterns used in the experiments, DIC analyses are performed on the generated reduced resolution images representing different values of the fill factor. In each case, the image corresponding the 5 mm rigid-body-translation is correlated with the reference position image. The DIC analysis is performed using the "MatchID-2D" [22] software. A square region of interest (size 800×800 pixels) located at the centre of the image is used in all correlations. The exact same correlation parameters are used for all three different speckle patterns in order to avoid any differences related to the correlation parameters. A subset size of 41×41 pixels with 20 pixels step size is used in the analysis. The size is chosen such that it will be suitable for the three different speckle patterns, especially speckle (c). Though this subset size is considered to be relatively large for speckle (b), since it has small dot size; however, there is no harm in using larger subset sizes for rigid-body-translation experiments [10, 23]. After obtaining the horizontal and vertical displacements U and V, the in-plane strains are calculated using 7×7 points strain window size.

## RESULTS AND DISCUSSION

When a surface undergoes a pure in-plane rigid-body-translation of a known magnitude, all points on that surface should theoretically have the same magnitude of translation and that makes all strain components to be zero. Using the rigid-body-translation experiments, the error in DIC measurements can be assessed using either the displacement or the strain results [12]. In order to assess the error of DIC measurements based on displacements, the displacement standard deviation for all points within the region of interest is calculated. Theoretically, all points on the surface should have the same displacement during rigid-body-translation and thus the standard deviation should be zero. Therefore, the value of the displacement standard deviation, usually reported in pixel units, represents the error in DIC displacement measurements. Alternatively, when the error assessment is based on strains, any strains obtained from the DIC analysis reflect an error in the results since the strain should in fact be zero. Similarly, the value of the strain standard deviation is a representation of the error in DIC strain measurements. By performing DIC analysis on the modified images having reduced fill factor and comparing the results with those of the images with 100% fill factor, the effect of fill factor on DIC measurement error may be realized. Both the displacement error and the strain error are calculated and compared for all the modified images for the three different speckle patterns.

As stated earlier, both speckle patterns (a) and (b) have high contrast and the main difference between them is the dot size. The reduced resolution images used for the analysis have a scale factor of 13 pixels/mm and that makes the dot diameters for speckle patterns (a) and (b) to be about 7 pixels and 3 pixels, respectively. On the other hand, speckle pattern (c) has relatively lower contrast. Figure 5 shows the DIC displacement and strain error results for the three speckle patters obtained using 100%, 50% and 25% fill factor images. For each of the image groups, the correlation is performed between the reference position image and the 5mm horizontal (*x* direction) translation image (the average translation was roughly about 64 pixels). By comparing the obtained DIC errors for speckle (a), it can be clearly seen that the fill factor has no effect on the error, since the images with different fill factors has almost identical error values. On the contrary, for speckles (b) and (c), the fill factor has some effect on both the displacement and strain errors. As can be seen from the figure, in general, speckle (b) shows the largest fill factor dependence with a percentage increase of 13.2% and 19.8% in the displacement and strain errors, respectively.

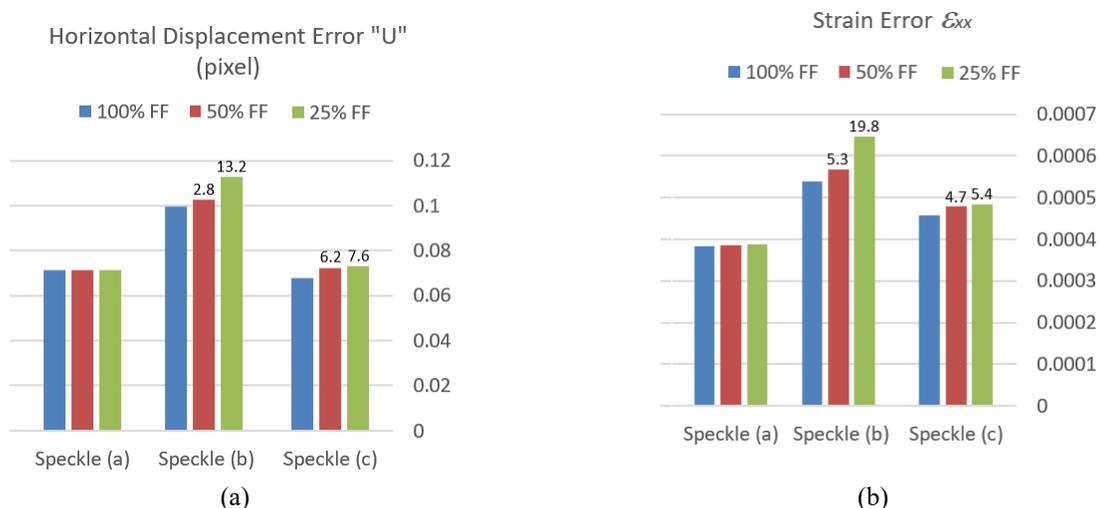

**Figure 5.** Effect of the fill factor on the DIC measurements error resulting from translation in the *x* direction (64 pixels) for the different speckle patterns; a) displacement error, b) strain error. (*for speckles "b" and "c" the percentage increase in error relative to the 100% fill factor images is shown on top of each column*)

In fact, the observed percentage increase in error for the 50% fill factor images is relatively smaller than what is expected based on the results reported in reference 12. However, a very interesting observation is revealed by inspecting the displacement and strain errors in the vertical direction (along the *y*-axis). Though the direction of translation in the experiments is along the *x*-axis (about 64 pixels), a small magnitude of translation of about 0.4 pixels in the vertical direction is detected by the DIC analysis. This very small amount of vertical translation is due to the imperfection of the alignment of camera's *x*-axis relative to the translating stage (such small error is unavoidable in experimental setups). Figure 6 shows the DIC vertical displacement and strain error results for the three speckle patters obtained using 100%, 50% and 25% fill factor images. It should be stressed here that these errors are mainly associated with the vertical translation (which is about 0.4 pixels). Similar to previous results shown in figure 5, the fill factor affects the error for speckle patterns (b) and (c) only. Interestingly enough, it can be seen from the figure that the fill factor has a very strong effect on the DIC measurement error. It can also be observed that the most significant increase in error occurs with the 25% fill factor images, while it is not that significant for the 50% fill factor images. Actually, this observation confirms that the increase in error is associated with the vertical translation component, since the main difference between the 50% and 25% fill factor images is the linear fill factor in the vertical direction (for the 50% fill factor images the linear fill factor in the vertical direction is 100% while for the 25% fill factor images the linear fill factor in the vertical direction is 50%). This clear and significant difference between the effects of fill factor on DIC measurement error seen in figures 5 and 6 is in fact associated with the magnitude of translation rather than the direction of translation. The results show that subpixel translations can lead to more significant error than larger translations of several pixels. This is likely to be due to the fact that for larger translations the optical aberrations of the lens come to play a more significant role and it can in some cases obscure the effect of the fill factor.

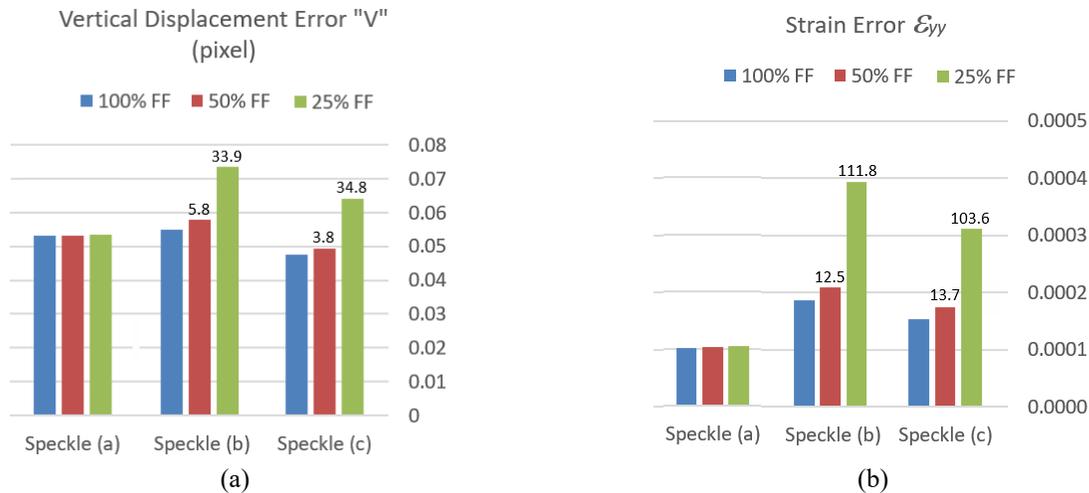

(a)                 (b)

**Figure 6.** Effect of the fill factor on the DIC measurements error resulting from translation in the *y* direction (0.4 pixel) for the different speckle patterns; a) displacement error, b) strain error. (*for speckles "b" and "c" the percentage increase in error relative to the 100% fill factor images is shown on top of each column*)

Both figures 5 and 6 show that the fill factor basically has no effect on speckle pattern (a) which consists of black dots on white background with 7 pixels dot diameter. Furthermore, speckle (a) consistently show the lowest magnitude of strain error in all cases. This observation clearly indicates that the use of such speckle patterns can evade the negative effects of low fill factor imaging sensors on the measurement accuracy of DIC and photogrammetric methods in general. On the other hand, speckle pattern (b) which is similar to speckle pattern (a) except for the dot size (the dot diameter is 3 pixels) shows the largest dependence on the fill factor. In fact, several studies have investigated the effect of speckle pattern on measurement accuracy and found that high contrast speckle patterns give the best measurement accuracy [1]. Furthermore, researchers also investigated the effect of the speckle pattern's dot size on measurement accuracy and concluded that an average dot diameter of 3 to 5 pixels gives the best results [24]. Similarly, in particle image velocimetry (PIV), which is a photogrammetric technique used in fluid mechanics applications, investigations on particle size have concluded that the recommended particle diameter is 3 pixels or larger [25]. The results presented in here, however, show that the use of speckle patterns with 3 pixels dot diameter can negatively affect the measurement accuracy when low fill factor cameras are used. But, interestingly enough, when the dot diameter is larger (about 7 pixels) the effect of fill factor on measurement accuracy can be mitigated.

**CONCLUSION**

The results of this study show that the FPA's fill factor can have a significant effect on the accuracy of DIC measurements. This effect is clearly dependent on the type of speckle pattern being used. The fill factor has a clear effect on measurement error for low contrast speckle patterns. On the other hand, for high contrast speckle patterns, black dots on white background, the fill factor has a diverse effect on measurement error depending on the dot size. For a dot diameter of 3 pixels, the fill factor strongly affects the measurement error but when the dot size is larger (about 7 pixels dot diameter) the fill factor has a very limited effect on the error. These results suggest that using high contrast speckle patterns with large enough dot size can eliminate the negative effect on measurement accuracy caused by using low fill factor imaging sensors. This also makes the use of costly image sensors fitted with micro-lens arrays to be not necessary in DIC experiments, provided that the illumination intensity is sufficient.

Additionally, the effect of fill factor on measurement error also depends on the magnitude of translation. The results show that the fill factor can affect the measurement error associated with small subpixel displacements more than large displacements of several pixels. This might be because the optical aberrations of the lens come to play a more significant role for larger displacements and that, in some cases, can obscure the effect of the fill factor. This observation will be subject to further investigation in future work.

**Acknowledgement**

The first author is pleased to acknowledge the financial support provided by the German Research Foundation (DFG) for funding his research visit to the Institute of Fluid Mechanics and Aerodynamics at Universität der Bundeswehr München.